# Nanocellulose Fragmentation Mechanisms and Inversion of Chirality from the Single Particle to the Cholesteric Phase

Gustav Nyström[1], Jozef Adamcik[1], Ivan Usov[2], Mario Arcari[1] and Raffaele Mezzenga[1,3*]

1. ETH Zurich, Department of Health Science & Technology, Schmelzbergstrasse 9, 8092 Zurich, Switzerland

2. Paul Scherrer Institute, 5232 Villigen PSI, Switzerland

3. ETH Zurich, Department of Materials, Wolfgang-Pauli-Strasse 10, 8093 Zurich, Switzerland

*Correspondence to: raffaele.mezzenga@hest.ethz.ch

**Understanding how nanostructure and nanomechanics influence physical material properties on the micro- and macroscale is an essential goal in soft condensed matter research. Mechanisms governing fragmentation and chirality inversion of filamentous colloids are of specific interest because of their critical role in load-bearing and self-organizing functionalities of soft nanomaterials. Here we provide a fundamental insight into the self-organization across several length scales of nanocellulose, an important bio-colloid system with wide-ranging applications as structural, insulating and functional material. Through a combined microscopic and statistical analysis of nanocellulose fibrils at the single particle level, we show how mechanically and chemically induced fragmentation proceed in this system. Moreover, by studying the bottom-up self-assembly of fragmented carboxylated cellulose nanofibrils into cholesteric liquid crystals, we show via direct microscopic observations, that the chirality is inverted from right-handed at the**

**nanofibril level to left-handed at the level of the liquid crystal phase. These results improve our fundamental understanding of nanocellulose and provide an important rationale for their application in colloidal systems, liquid crystals and nanomaterials.**

The increasing global population and improved living standards demand a more efficient use of available resources and energy[1]. Improving the use of sustainable resources without sacrificing the final materials performance is therefore critical[2]. Nanocellulose represents an important class of sustainable bio-colloids with great promise for use in new high strength[3], energy storage[4] and insulation[5] nanomaterials. To understand and identify nanostructural and nanomechanical details, including the processing-structure relationship, is essential for advancing the understanding of these bio-colloids and their application in new materials. Specifically, investigating how these nanoparticles kink[6] and break is of fundamental interest for their load bearing[7], gel forming[8] and self-organizing[9–11] functionalities.

It is well known that mechanical energy (sonication) breaks cellulose nanofibrils (CNF)[7] and cellulose nanocrystals (CNC)[12] resulting in shorter particles. It is also well known that acid hydrolyzes cellulose resulting in short CNC particles[13]. However, the mechanisms behind these nanocellulose fragmentation processes remains, to the best of our knowledge, largely unknown. In this paper we set out to answer the open questions whether or not fibril breakage occurs at positions of kinks or along rigid segments, if kinks can be introduced mechanically[7,14] and if acid hydrolysis occurs, as previously suggested[15–18], preferentially at positions of kinks.

Nanocellulose, mainly in the form of sulfate half-ester modified CNC, is known to self-assemble into cholesteric liquid crystalline phases[9]. This discovery has opened up an entire field of research on sulfated nanocellulose cholesterics[10,19–24]. While the results are impressive, the detailed mechanism behind the formation of the cholesteric phase remains unresolved[25]. One hypothesis is that the cholesteric phase forms as a result of the twisted shape of the cellulose particle[26]. Unfortunately, the nanostructure of sulfated CNC is highly irregular and the chirality of single CNC particles is difficult to unambiguously determine. Twisting has, however, previously been observed on thicker cellulose fibril aggregates,[26,27] but until recently, the chirality of the single cellulose nanoparticle remained unknown. Recent results revealed that single carboxylated nanocellulose particles have a well-defined right-handed chirality[14], opening up the unprecedented possibility to shed light on how chirality transfers across length scales in the nanocellulose system. Understanding the underlying mechanisms of chirality and how it transfers between different hierarchical levels is a fundamental research challenge with important implications spanning from self-assembled biological nanomaterials to biochemistry, biophysics and nanotechnology.

In the present study, we chose carboxylated CNF as model system for the following three reasons. First, it allows us to study nanocellulose fragmentation mechanisms and cholesteric self-assembly behavior on a single well-defined system. Second, this approach gives us, discussed in the first part of the paper, the possibility to control the detailed sample nanostructure before and after fragmentation making it possible to draw conclusions about how fibrils break as a result of mechanical and acid treatment. Third, the carboxylated CNF and CNC have a well-defined right-handed chirality. This gives us, disclosed in the second part of the paper, the possibility to explicitly study how the chirality of the single particle transfers to the chirality of the cholesteric liquid crystal phase.

**RESULTS**

**Mechanically induced fragmentation of carboxylated CNF.** To investigate the effects of mechanical treatment (sonication) on the nanostructure of single cellulose nanofibrils (CNF) atomic force microscopy (AFM) imaging was performed on samples collected after different sonication times, see **Fig. 1a–c**. Directly evident from these images are the right-handed chirality (bottom left corner of panel a) of the ~2 nm thin fibrils, in agreement with our earlier report[14], and an apparent shortening of the fibrils as a result of the increased sonication time. To gain more insight into the nanostructural changes of the fibrils, statistical image analysis was performed using our own-developed open-source code FiberApp[28], allowing collection and analysis of the fibril contours, including the number and position of the kinks (defined using white square masks), see Fig. 1d. Using the positions of the masks, fibril contours could be artificially split into shorter rigid contours, called segments. The tracked data from 2628 fibrils were then used to extract three arithmetically averaged parameters for each sonication time, namely the contour length of the whole fibril, the inverse linear kink density (equivalent to the average distance between kinks) and the segment length. As a result of the sonication treatment, we propose that three different events could occur, either: i) a new kink could appear, ii) the fibril breaks at the position of a kink or iii) the fibril breaks along a rigid segment. These possible events and their hypothetical, ensued influence on the parameters of interest are summarized in Fig. 1e. For instance, breakage at the position of a kink would lead to a decreased average contour length, an increased average inverse kink density (this event is statistically equivalent to repairing a kink) and an unchanged average segment length. In Fig. 1f the averages of the contour length, the inverse kink density and the segment length are plotted for four different sonication times (0 s, 60 s, 240 s and 420 s). The data, found to be largely invariant as to whether or not fibrils only containing kinks were included in the analysis, shows that the average contour length significantly decreases (unpaired t-test, p-value (P) < 0.0001) with increasing sonication times. This is in agreement with previous results[7] and the visual comparison in Fig. 1a–c. Within the first 60 seconds, the rapid decrease in the average contour length is followed by a decrease in the average segment length (P < 0.0001), but no statistically significant change in the inverse kink density. In combination, this would indicate that the initial mechanical breakage occurs along a segment and not, as one might assume, at a position of a kink. We note, however, that the breakage along rigid segments can be seen as a superposition of the other two events, meaning that we can also not discard the possibility that this breakage occurs by first introducing a new transient kink followed by breakage at the same kink position. For longer sonication times the data indicates further breaking along the segments, based on the significant decrease of both the contour length and the segment length (P < 0.0001). Although no change was found during the initial 60 s of sonication, for longer sonication times there is also a significant (P < 0.0001) decrease in the inverse kink density. While in general all three events are expected to occur simultaneously, only introduction of new kinks can explain a decrease in inverse kink density. These data are therefore conclusively inferring, in line with previous results[7,14], that the observed kinks are a result of the mechanical treatment of the sample rather than the presence of amorphous regions along the fibril contour.

Comparing the relative values for the arithmetic and the weight average of the segment length before and after sonication treatment, a 47% decrease is found for the number averaged mean compared to a 65% decrease obtained for the weight averaged mean. This difference supports the interpretation that rigid segments are broken and highlights that longer segments are preferentially broken, as would be expected based on energy considerations where the probability for breakage increase with L for sufficiently large energies ($P \sim e^{-(E/kT)} * L$ for $E > E_c$).

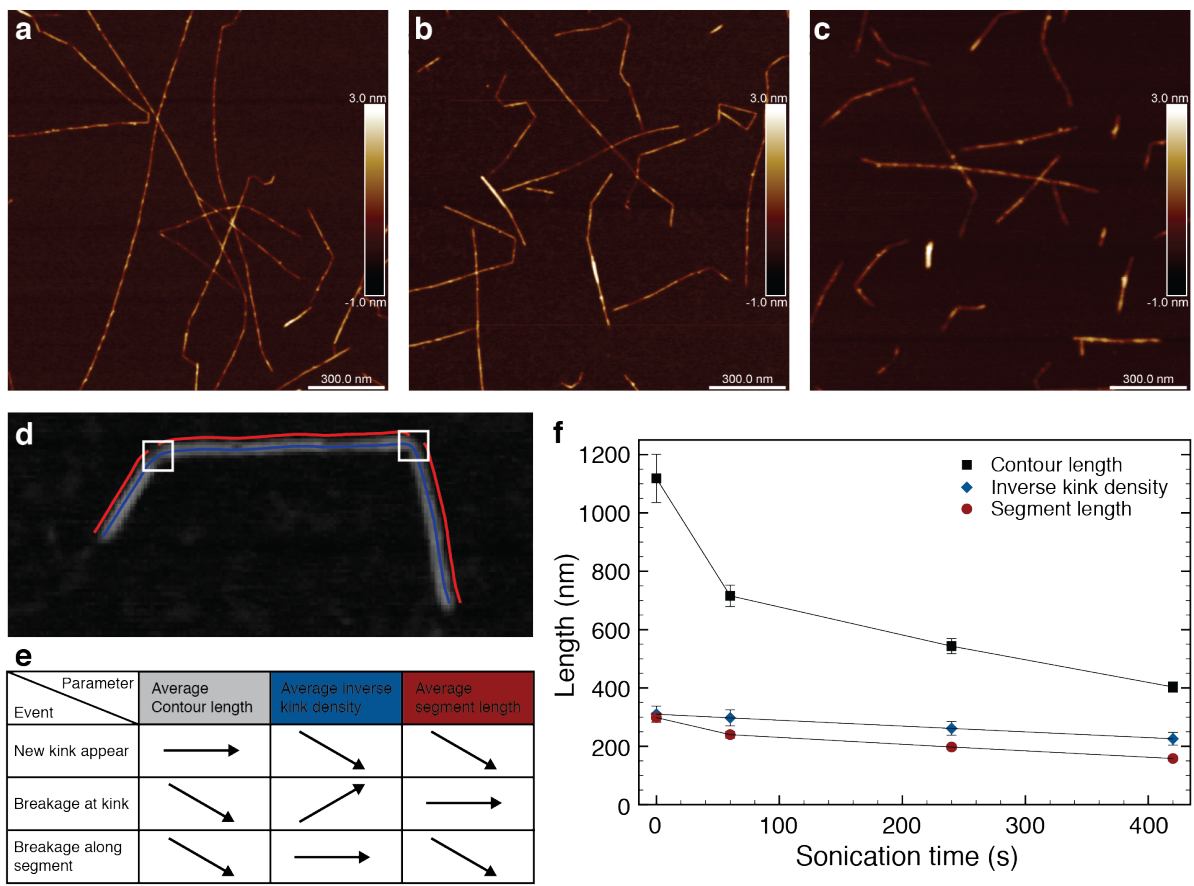

**Figure 1. Influence of sonication on carboxylated cellulose nanofibril (CNF) nanostructure.** Tapping mode AFM-images of cellulose nanofibrils before (**a**) and after being subjected to 60 s (**b**) and 240 s (**c**) of sonication treatment. (**d**) Magnified section from one of the AFM images highlighting the tracking of one cellulose nanofibril including its tracked contour (blue line), masks to treat the kink region (white squares) and splitting of the contour in the middle of the masks into three segments (red lines). (**e**) Table summarizing the possible events that could occur to cellulose nanofibrils during sonication (column 1) and its expected influence on the nanofibril parameters of the nanofibrils (columns 2–4). (**f**) Plot of the arithmetic averages of the contour length (black squares), the inverse kink density (blue diamonds) and the segment length (red circles) as a function of sonication time with 95% confidence interval error bars.

**Hydrochloric acid induced fragmentation of carboxylated CNF.** Besides mechanical treatment, it is also well known that mineral acids can cause hydrolysis of cellulose, resulting in a cleaving of the cellulose chains[15] as well as a shortening of the cellulose nanofibrils[29]. A commonly held belief is that the hydrolysis of cellulose nanofibrils occurs preferentially at positions of disordered cellulose chains, ascribed to regions at which the fibrils kink [17]. In order to test whether or not this hypothesis holds true, we first envision three different possible outcomes of the hydrolysis of kinked cellulose nanofibrils as schematically indicated in **Fig. 2a**. The hydrolysis would occur either i) only at kinks, ii) preferentially at kinks or iii) at random positions along the nanofibril contour. To study the effects of acid hydrolysis, kinked fibrils were imaged using AFM before and after hydrolysis with hydrochloric acid (see Methods for details), as shown in Fig. 2b–c. As the images show, only in rare cases are there kinks still present in the hydrolyzed sample. This effectively rules out the scenario of a random hydrolysis process, but is compatible with both a hydrolysis process occurring only or preferentially at kinks. To distinguish between these two scenarios the length distribution of hydrolyzed particles is compared to the distribution of rigid segments in the non-hydrolyzed sample, see Fig. 2d. From this comparison it is evident that the distribution after hydrolysis is shifted to shorter lengths compared to the original rigid segment distribution. Moreover, a large amount of particles with a length shorter than the minimum distance between kinks in the non-hydrolyzed sample is observed. Combined with the relative absence of kinks, we conclude that the acid hydrolysis occurs according to mechanism ii) preferentially but also in between kinks. With short, mainly unkinked, right-handed chiral cellulose nanoparticles at hand, we now turn to the question of how the chirality is transferred from the single particle to the liquid crystal phase.

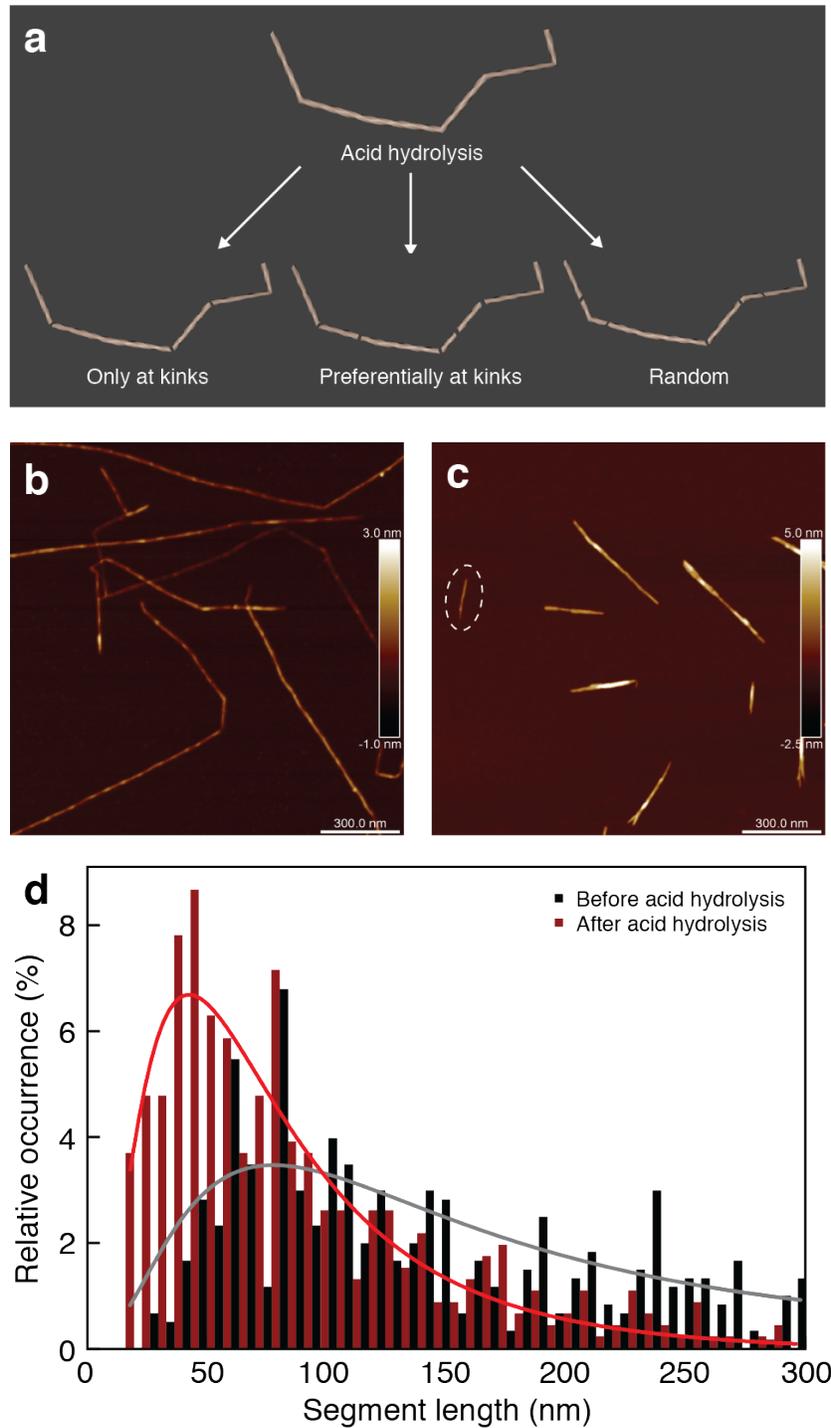

**Figure 2. Hydrolysis of carboxylated cellulose nanofibrils (CNF) into carboxylated cellulose nanocrystals (CNC).** (**a**) Schematic describing possible results of cutting the cellulose nanofibril into shorter cellulose nanocrystals. AFM images of cellulose nanofibrils before (**b**) and after (**c**) acid hydrolysis. Only segments with the same thickness (indicated by the dashed white line) as the untreated fibrils were included in the analysis. (**d**) Segment length distribution before (black bars) and after (red bars) acid hydrolysis. The data has been binned equally and is plotted around the center of each bin position (the red data shifted to the left) for clarity.

**Nematic and cholesteric phases of carboxylated CNC.** With an increasing concentration, the system of the hydrolyzed short cellulose fibrils, CNC, undergoes an isotropic-nematic phase transition[9,12,20,26,30] as indicated schematically in **Fig. 3a**. For the present carboxylated CNC system the aspect ratio is ~95 meaning that the Onsager formalism[31] can be applied quantitatively. Taking also the charges on the particles into account[32], the isotropic-nematic phase transition is given by:

$$\phi_{IN} = 6 \frac{D^2}{D_{eff} L} \qquad (1)$$

where D is the diameter, $D_{eff}$ the effective diameter and L is the length of the CNC (see Methods for details). For the effective linear surface charge density $\nu_{eff}$ = 0.26 e/nm, <D> = 3.8 nm, $D_{eff}$ = 17.9 nm and <L> = 366 nm a transition at a volume fraction of $\phi_{IN,pred.}$ = 0.0132 is predicted. Considering the polydispersity of the particles and the approximations used to estimate their effective charge density, this is in reasonably good agreement with the observed transition at ~8 g/L (Fig. 3b) or $\phi_{IN,exp.}$ = 0.0053. This concentration is also in perfect agreement with previous observations of ordering of similar particles using small angle neutron scattering[11]. Under the present experimental conditions, where the system is close to the isotropic-nematic phase transition, both phases coexist and the phase separation is proceeding *via* nucleation of small nematic liquid crystal tactoids that grow in size over time (see Fig. 3b–e), from an initially isotropic metastable dispersion. At a certain critical tactoid volume[33], the surface energy constraints on the structure within the tactoids are sufficiently relaxed, allowing a cholesteric phase to develop (see Fig. 3e). Over time, these cholesteric tactoids grow and eventually merge[30] to a bulk cholesteric phase (see Fig. 3f).

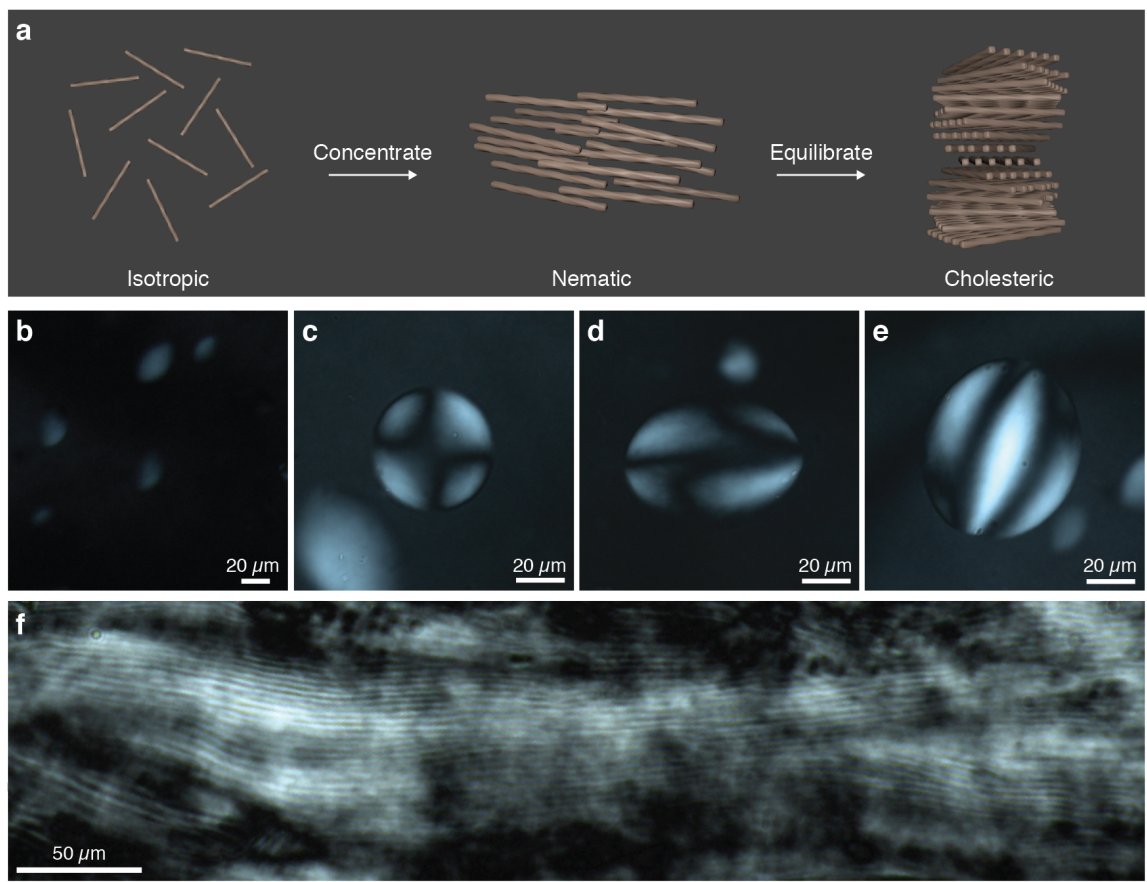

**Figure 3. Carboxylated CNC assembly in nematic and cholesteric phases.** (**a**) Schematic describing the entropic ordering of rod-like particles in nematic and cholesteric liquid crystal phases. Nucleation of small nematic tactoids out of the surrounding isotropic phase (**b**) that grow in size with time (**c**, **d**) eventually displaying the cholesteric liquid crystalline phase (**e**). (**f**) Bulk cholesteric phase observed after all initially metastable isotropic phase has been converted into a cholesteric phase.

**Transfer of chirality from particle to cholesteric phase for carboxylated CNC.** Although the formation of the cholesteric liquid crystal phase is well known for nanocellulose[9], the general underlying mechanism behind the formation of the cholesteric phase is still not completely resolved[25]. Chirality on the single particle level is a necessary but not sufficient condition to develop a cholesteric phase since also the thermodynamic state of the system[34] and molecular scale variations play a role. For instance filamentous bacteriophage Pf1 and fd viruses both have a right-handed chirality, but Pf1 displays only a nematic phase whereas fd virus develops a cholesteric phase[35]. Furthermore, a single point mutation of the major coat protein was found to change the chirality of the cholesteric phase from left-handed (for *fd wt* virus) to right-handed (for *fd* Y21M)[36]. Our previous results[14] conclusively showed that the single chirality of carboxylated nanocellulose particles is right-handed. This is very valuable information, which is not available for the more commonly studied cholesteric system of sulfated cellulose nanocrystals, and which allows us, for the first time, to study how the chirality is transferred from the single nanocellulose particle to the cholesteric phase. For this purpose we designed an experimental setup based on polarized optical microscopy where the handedness of the cholesteric tactoids could be directly determined, see Fig. 4a. By orienting the sample cuvette at a 45 degree angle to the polarizer and analyzer (x- and y-axis in Fig. 4a), tactoids with their cholesteric axis parallel to the y'-axis will display a maximum cholesteric band intensity. Rotating the cuvette around the y'-axis will therefore also cause a rotation of y'-aligned tactoids around their cholesteric axis. This tactoid rotation allows us to determine the handedness of the cholesteric phase by following the rotation-induced movement of the cholesteric bands. When rotating the sample clockwise (right-handed), we observe an upward movement of the maximum intensity bands (Fig. 4b). This observation is in agreement with the right-handed rotation of the left-handed schematic helix in Fig. 4c. For the present system we accordingly observe a left-handed chirality for the cholesteric droplet and further, during a rotation of ~90° the bands of maximum intensity shift, as expected, by a distance corresponding to P/4, where P is the cholesteric pitch. Remembering the right-handed chirality of the cellulose nanofibrils, a chirality inversion to left-handedness is observed for the nanocellulose cholesteric liquid crystalline phase. This follows a similar mechanism of chirality inversion compared to DNA[37] and virus[38] and similar but opposite to that of amyloid fibrils[33] where left-handed particles develop right-handed cholesteric phases. Additionally, the effect of the temperature on the equilibrium bulk pitch of the carboxylated CNC cholesteric phase was investigated, see Fig. 4d. Unlike the systems of virus[35] and DNA[39] that show an increase and decrease respectively of the pitch with increasing temperature, but in accordance with previous results on sulfated CNC[40], the carboxylated CNC shows a cholesteric pitch that is not influenced by the change in temperature, at least in the range explored. This suggests that the cholesteric phase develops mainly as a result of entropic hard-core interactions between the chiral rod-like particles[26] in line with the model of Straley[41], allowing a chiral inversion for sufficiently large (> 45°) particle twist angles[38] (Fig. 4e). While it is difficult to define an exact twist angle for the present system, it is reasonable, based on the AFM images, to assume large values for the particle twist angle. The large twist angle may also, in part, explain why a defined twist and handedness are difficult to be assigned for the shorter particle sulfated CNC system.

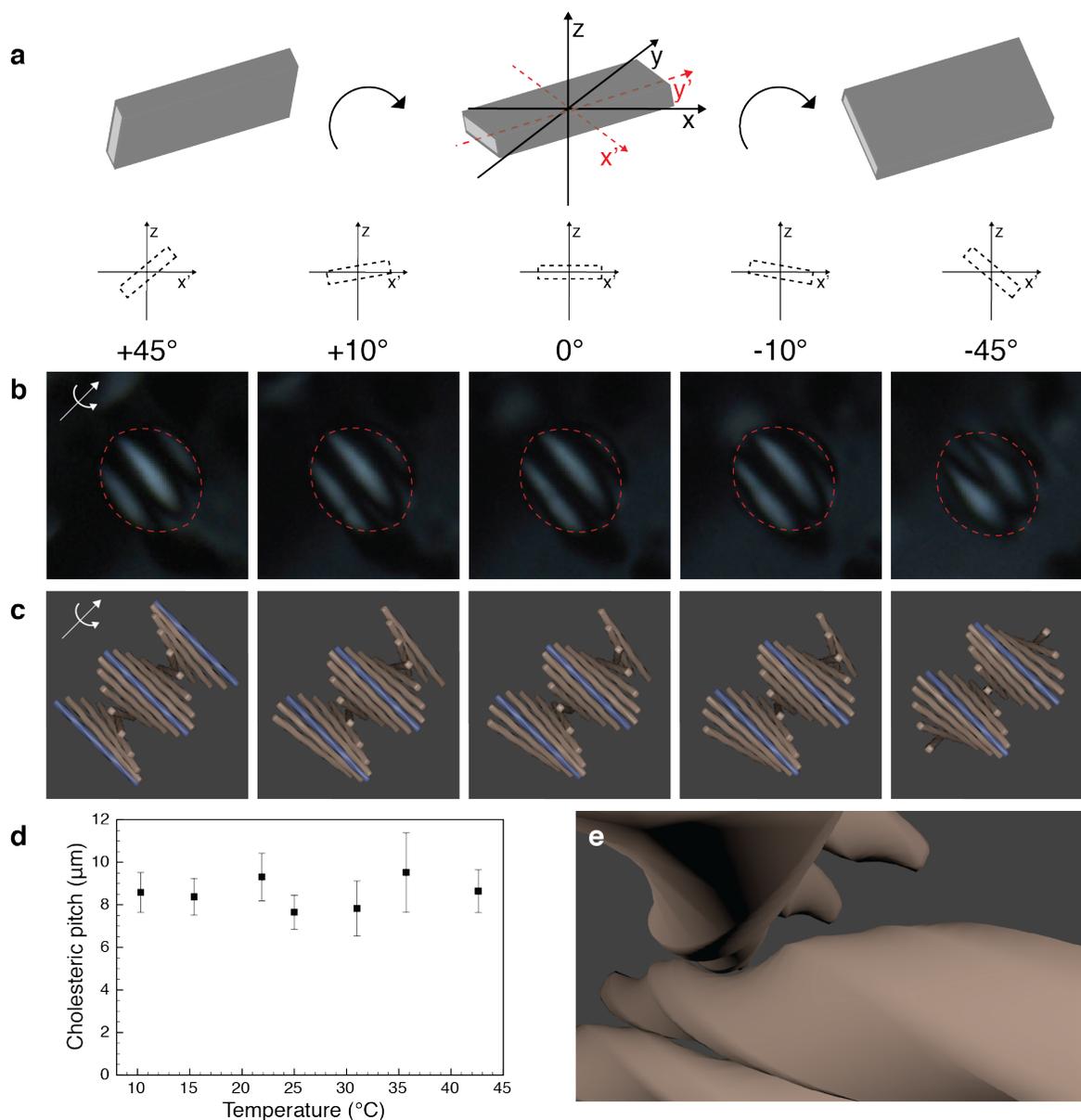

**Figure 4. Determination of handedness and temperature dependence of the cholesteric phase.** (**a**) Schematic describing the right-handed rotation of the sample around its long axis (y') and the resulting rotation angles in the x'-z plane. (**b**) Corresponding rotation of a cholesteric tactoid, and translation of bands of maximum intensity, as observed between crossed polarizers in the optical microscope. (**c**) Comparing schematic of a right-handed rotation of a left-handed rod-helix. The blue rods correspond to positions on the helix with maximum intensity. (**d**) Influence of the bulk cholesteric pitch as a function of temperature. The error bars represent the standard deviation of 5 independent measurement of the pitch each averaged over 8–10 cholesteric bands. (**e**) Zoomed schematic illustrating the steric (left-handed) interaction of the cholesteric phase between two adjacent right-handed rods.

**Chirality of solid-state carboxylated CNC films.** As a further demonstration of the (bulk) chirality of the system, dry films, cut along an angle from the film surface, were imaged with AFM as shown in Fig. 5. These images, for both the carboxylated CNC of the present study (Fig. 5a) as well as the sulfated CNC added as a control sample (Fig. 5b), show an arc pattern characteristic of the cholesteric phase. The carboxylated CNC sample, although less regular, shows the same arcing behavior as the sulfated CNC. This counter-clockwise rotation of the local director, while moving from the top (upper left corner) to the bottom (lower right corner) of the film corresponds to a left-handed cholesteric phase, and is in agreement with both the present direct observation of single cholesteric tactoids using the polarized light microscope and previous results based on electron microscopy[10].

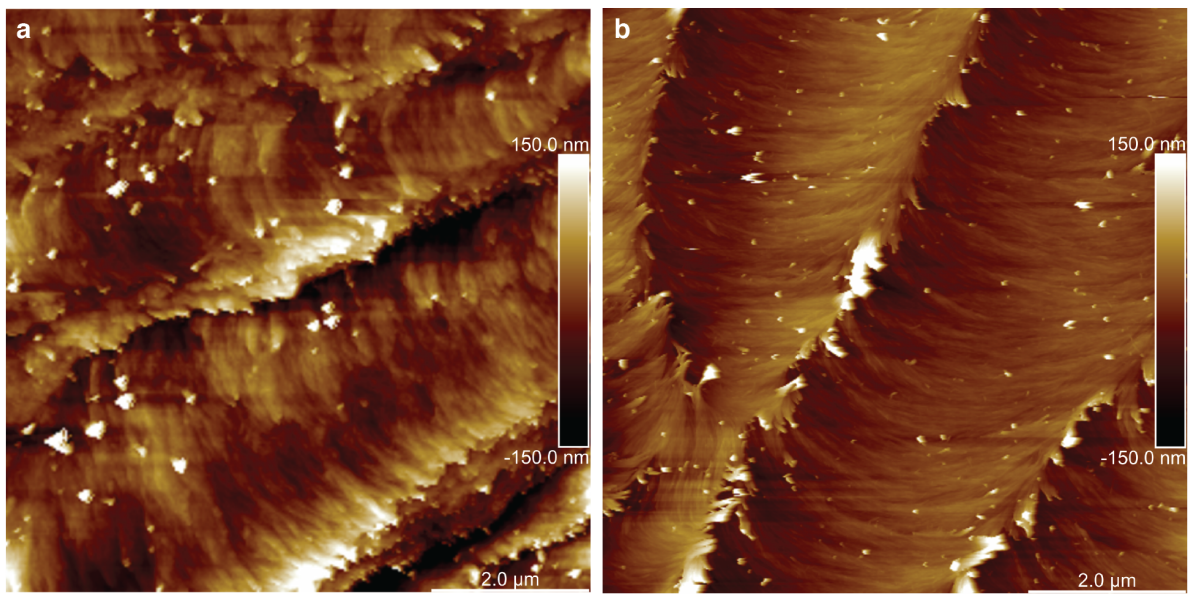

**Figure 5. AFM of solid-state CNC films.** Scans across inclined surfaces in (**a**) CNC-COOH and (**b**) CNC-SO3H films oriented with the top left corner away from the observer showing arc patterns indicative of a left-handed cholesteric phase.

In summary, we have studied the fragmentation of nanocellulose using sonication and acid hydrolysis and found that there is a rapid initial breakage due to sonication that occurs at rigid segments along the fibril contour. For longer sonication times, the data shows a continued fracturing of the fibrils at the rigid segments as well as an increased density of kinks inferring that the presence of kinks is – at least partially –a result of the mechanical sample treatment. The results based on acid hydrolysis demonstrated that the cutting of fibrils occurs preferentially at kinks but also at random position in between kinks, creating a lower particle length distribution than what would be expected based on the original rigid segment length distribution only. By direct microscopic observation, we finally showed that right-handed carboxylated CNC self-organize into cholesteric tactoids and films of a left-handed handedness, demonstrating an inversion of the chirality between the single particle and the liquid crystal phase level. Not only do these results provide important structural information and an improved understanding of nanocellulose on the fundamental level, but they also offer a critical step forward in their application in new nanomaterials and nanotechnologies.

**ACKNOWLEDGEMENTS**

G.N. acknowledges funding from the Gunnar Sundblad Research Foundation and from the Swiss National Science Foundation Ambizione Grant No. PZ00P2_168023/1.

**AUTHOR CONTRIBUTIONS**

G.N., I.U., J. A. and R.M. designed the study. G.N., J.A. and M.A. performed the experiments. G.N., I.U. and R.M. analyzed data. G.N., M.A. and R.M. wrote the paper.

**METHODS**

**Preparation of carboxylated cellulose nanofibrils (CNF).** Carboxylated cellulose nanofibrils were prepared from never-dried softwood pulp using TEMPO-mediated oxidation[42] combined with mechanical treatment (ultrasonication) and centrifugation to remove larger fibril aggregates. The total amount of charge (0.67 mmol/g) was measured using conductometric titration on the oxidized pulp. The oxidized pulp was then dispersed at a concentration of 1 g/L and cellulose nanofibrils were extracted using probe ultrasonication for 2 minutes (Hielscher UP200S, operated at 200 W and 20% amplitude setting) followed by centrifugation at 12 000 g for 30 minutes.

**Preparation of carboxylated CNF for sonication experiment.** A common 1 g/L stock dispersion of CNF with 0.67 mmol/g charge density was split into three identical 20 mL glass vials each filled with 5 mL of dispersion. Probe ultrasonication (Hielscher UP200S, operated at 200 W and 20% amplitude setting) was thereafter performed using a 7 mm probe immersed in the middle of the dispersions for 60, 240 and 420 s respectively. To avoid heating the vials, these were placed in an ice bath during the sonication treatment.

**Preparation of carboxylated cellulose nanocrystals (CNC).** Cellulose nanocrystals were prepared by heating 80 mL of 0.5 g/L of the carboxylated CNF dispersion with a charge density of 1.5 mmol/g at 105 °C and 2.5 M HCl under stirring at 200 rpm and reflux conditions for 3 hours. The sample was collected by centrifugation (30 minutes and 12 000 g) and dialyzed against MQ-water for 3 days after which a final pH of 4.6 was reached. To disperse the CNC, the sample was sonicated for 2 min at 40% amplitude with a 7 mm probe followed by centrifugation (30 minutes and 12 000 g). The CNC concentration was measured gravimetrically and adjusted using reverse osmosis against a 10 wt% PEG 35 kDa solution. The samples that were used for analysis of the particle length distribution after hydrolysis were collected after dialysis and only briefly sonicated for 20 s at 20% amplitude settings to assist the dispersion of single CNC particles.

**Characterization by optical microscopy.** The samples for optical microscopy were prepared in square cross section (0.20 x 4.00 mm) glass cuvettes (VitroTubes, Vitrocom) and sealed with epoxy glue. The cuvettes were stored horizontally until the epoxy glue had hardened completely, at least 30 minutes. After that they were labeled and stored vertically at room temperature.

Polarized optical microscopy was performed using a Zeiss Axioscope 2 microscope using 5x and 10x air objectives to image the samples between crossed linear polarizers. An in-house made rotation stage was used to rotate the sample in the horizontal sample plane and around the long axis of the sample. To determine the handedness of the cholesteric pitch, the cholesteric tactoid was oriented at a 45-degree angle to the crossed polarizers with the cholesteric bands at a right angle to the polarization plane. The rotation was then performed along the helical axis of the sample and the resulting movement of the position of the cholesteric bands was used to determine the handedness of the cholesteric phase. If the cholesteric tactoid is rotated in a right-handed fashion, a movement of the bands in the opposite direction of the helix corresponds to a right-handed cholesteric phase, while an opposite motion corresponds to a left-handed cholesteric. It is at this point very important to take the transformation between the sample and the image plane in the microscope into account. In our setup there is a 180-degree rotation between the sample and the image plane leaving the observation of the handedness unchanged.

**Characterization with atomic force microscopy.** A droplet of 0.01 g/L dispersion was deposited on freshly cleaved mica and allowed to adsorb for 60 s before rinsing with MQ-water and drying with pressurized air. AFM measurements were performed using a MultiMode VIII Scanning Probe Microscope (Bruker, USA) covered with an acoustic hood to minimize vibrational noise. AFM images were acquired continuously in the tapping mode under ambient conditions using commercial cantilevers (Bruker).

**Statistical image analysis**. The high resolution (5120 x 5120 pixels) AFM images were analyzed using FiberApp[28] which allows each fibril to be represented by its contour – a sequence of points connected with straight segments that are positioned along the fibril maximum height on an AFM image. All contours acquired in this study have a constant distance between projections of these points on the image plane, which is the step size s = 1 pixel. The regions around kinks were treated using the previously described mask method,[14] which allows to set more flexible tracking parameters in the vicinity of kinks and more stiff parameters along rigid segments.

For the contour length, rigid segment and inverse kink density analysis after sonication was run on a total of 2628 fibrils, 657 for each sample set. Both fibrils with and without kinks were included in the analysis. To avoid any bias, the relative number of unkinked fibrils was kept within 14–20 % for all sample sets. For the rigid segment distributions before and after acid hydrolysis a total of 662 fibrils were analyzed. 200 fibrils before hydrolysis were analyzed containing a total number of 441 rigid segments. These were compared to the length distribution of 462 fibrils after hydrolysis. While doing the analysis care was taken to only include particles with a height comparable to that of the original particle.

To estimate the aspect ratio of the CNC sample forming the nematic and cholesteric phase, 407 particles were analyzed, and the arithmetically averaged lengths (366 nm) and heights (3.8 nm) were used giving an aspect ratio of ~95.

The rigid segment and contour length distributions were fitted using a log-normal probability density function $f(L; \mu, \sigma)$:

$$f(L;\mu,\sigma) = \frac{A}{L\sigma\sqrt{2\pi}} e^{-\frac{(\ln(L)-\mu)^2}{2\sigma^2}}$$

where L is either the rigid segment or the total contour length, μ and σ are the mean value and the standard distribution of the length's natural logarithm, respectively, and A is a distribution normalizing constant.